\documentclass[cit]{PoS}
\usepackage[authoryear]{natbib}
\bibpunct{(}{)}{;}{a}{}{,}

\title{
 Three-dimensional Tomography of the Galactic and Extragalactic Magnetoionic Medium
with the SKA
}
\ShortTitle{3D Tomography of the Magnetoionic Medium} 
\author{\speaker{J. L. Han}$^1$, W. van Straten$^2$, T. J. W. Lazio$^3$,
  A. Deller$^4$, C. Sobey$^4$, J. Xu$^1$, D.~Schnitzeler$^5$,
  H. Imai$^6$, S. Chatterjee$^7$, J.-P. Macquart$^8$, M. Kramer$^5$,
  J. Cordes$^7$
\\
$^1$National Astronomical Observatories, Chinese Academy of Sciences, Beijing 100012, China; 
$^2$Swinburne University of Technology, Australia; 
$^3$Jet Propulsion Laboratory, California Institute of Technology, USA;
$^4$ASTRON, The Netherlands; 
$^5$MPIfR, Germany;
$^6$Kagoshima University, Japan;
$^7$Cornell University, USA;
$^8$Curtin University, Australia.
\\ 
  E-mail: \email{hjl@bao.ac.cn}
       }

\abstract{
The magneto-ionic structures of the interstellar medium of the Milky
Way and the intergalactic medium are still poorly understood,
especially at distances larger than a few kiloparsecs from the Sun.
The three-dimensional (3D) structure of the Galactic magnetic field and
electron density distribution may be probed through observations of
radio pulsars, primarily owing to their compact nature, high
velocities, and highly-polarized short-duration radio pulses.
Phase 1 of the SKA, i.e.\, SKA1, will increase the known pulsar
population by an order of magnitude, and the full SKA, i.e.\, SKA2, will
discover pulsars in the most distant regions of our Galaxy.
SKA1-VLBI will produce model-independent distances to a large number
of pulsars, and wide-band polarization observations by SKA1-LOW and
SKA1-MID will yield high precision dispersion measure,
scattering measure, and rotation measure estimates along thousands of
lines of sight. When combined, these observations will enable detailed
tomography of the large-scale magneto-ionic structure of both the
Galactic disk and the Galactic halo.
Turbulence in the interstellar medium can be studied through the
variations of these observables and the dynamic spectra of pulsar flux
densities. SKA1-LOW and SKA1-MID will monitor interstellar
  weather and produce sensitive dynamic and secondary spectra of
pulsar scintillation, which can be used to make speckle images of the
ISM, study turbulence on scales between $\sim10^8$ and $10^{13}$ m,
and probe pulsar emission regions on scales down to $\sim$10 km.
In addition, extragalactic pulsars or fast radio bursts to be
discovered by SKA1 and SKA2 can be used to probe the electron
density distribution and magnetic fields in the intergalactic medium
beyond the Milky Way.}

\FullConference{Advancing Astrophysics with the Square Kilometre
  Array\\ 8-13 June 2014\\ Giardini Naxos, Italy}

\begin{document}

\section{Introduction} 

Our Milky Way consists of the Galactic disk, which has a few spiral
arms, the extended Galactic halo, the Galactic bulge, and a central
bar. However, a detailed picture of the Milky Way is not yet clear.
Spiral structures have been observed only in the closest half of our
Milky Way, and the pitch angles of only a few arm segments have
been determined \citep[see][]{hh14}. Embedded in these large-scale
structures, the interstellar medium (ISM) consists of diffuse neutral
hydrogen (H\,\textsc{i}) gas and H\,\textsc{i} clouds, higher density
regions of molecular clouds, diffuse ionized gas (H\,\textsc{ii})
and extended or compact H\,\textsc{ii} regions. In turn, the ISM is
permeated by magnetic fields, and the magnetized interstellar plasma
is known as the magnetoionic medium.  Although the magnetoionic medium
occupies a significant fraction ($\sim 0.2$) of the volume of the ISM,
we have only limited knowledge of its distribution in the disk and
halo. 
A considerable amount of energy is required to keep the ISM ionized,
roughly 15--20\% of the luminosity of all O/B stars in the Galaxy. 
Accordingly, ionized gas traces the energy input into the ISM, which
is dominated by supernova remnants and the stellar winds of bright
young stars. Therefore, in the Galactic disk, more ionized gas and
more disturbed small-scale structures are expected to be associated
with the spiral arms (as opposed to the interarm regions). However,
the origin of hot gas in the Galactic halo, where very few hot stars
exist, is not clear.

Magnetic fields are ubiquitous throughout the Universe. The magnetic
fields in our Galaxy play a crucial role in numerous astrophysical
processes: they affect the propagation of cosmic rays, impact the
evolution of molecular clouds and star formation, and facilitate
the transport of heat, angular momentum and energy. Measurements of
the magnetic structure of the Milky Way are sparse, especially at
distances greater than several kiloparsecs from the Sun and on length
scales shorter than tens of parsecs. Zeeman splitting measurements
probe the in situ magnetic fields of small-scale star formation
regions, which are evidently related to the large scale structure of
the Galactic magnetic field (\citealp{hz07}; see also \citealp{g+14} 
and \citealp{rob14} in this volume).


Faraday rotation measurements probe the integrated magnetic field
along the line of sight to pulsars and extragalactic radio sources.
Because pulsars are distributed throughout the Milky Way, their
compact nature and short-duration, highly-polarized radio pulses make
them ideal probes of the three-dimensional structure of the diffuse
magnetoionic medium in our Galaxy. Extraction of the common
contribution to Faraday rotation measures of extragalactic radio
sources yields the total Faraday rotation due to the Galactic
magnetoionic medium (\citealp{xh14a,xh14b,ha+14} in this volume).

\begin{figure}[hbt]
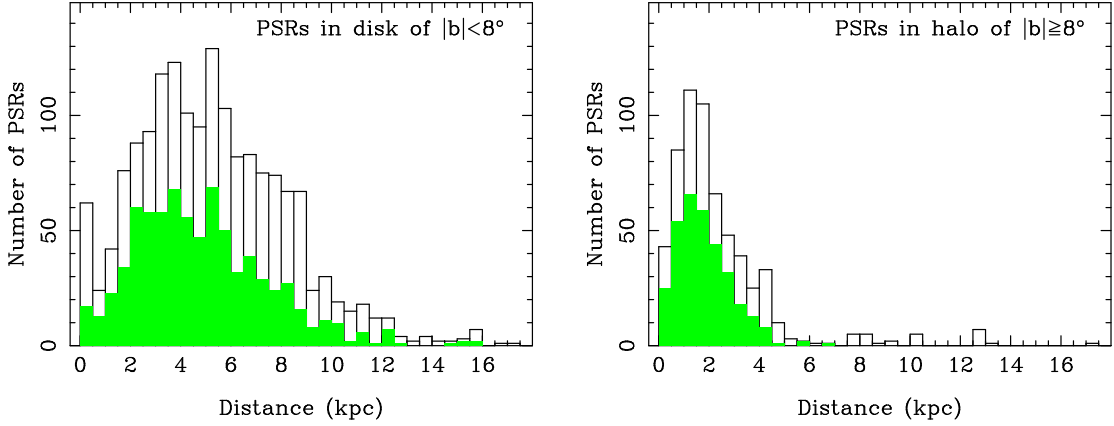

\centering
\includegraphics[angle=270,width=70mm]{fig1a.ps} \hspace{5mm}
\includegraphics[angle=270,width=70mm]{fig1b.ps}
\caption{Distance distribution of known pulsars in the Galactic disk
  ($|b|<8^{\circ}$) and in the Galactic halo $|b| \ge 8^{\circ}$. The
  green histograms correspond to pulsars with available RMs. }
\label{psrdist}
\end{figure}

To date, 2300 pulsars have been discovered using single dish
telescopes \citep[see the updated catalog of][]{mhth05}.  Most
of the known pulsars are in the Galactic disk, and about 20\% of them
are in the Galactic halo (see Figure \ref{psrdist}). The distribution
of pulsars at low Galactic latitudes occupies about half of the
Galactic disk; at high latitudes, the distribution is limited to about
4 kpc from the Sun. A large spatial volume of the Milky Way is left
for future pulsar discovery by the next generation of radio telescopes
with extremely high sensitivity, such as FAST \citep{nan11},
  LOFAR \citep{cvh+14} and the SKA \citep{kbk+14}.

According to the current SKA1 Baseline Design \citep{dtm+13},
Phase 1 of the SKA will consist of three arrays: a low-frequency array
(SKA1-LOW), a mid-frequency array (SKA1-MID), and a wide-field survey
array (SKA1-SUR). SKA1-LOW will be able to observe the
$-67^{\circ}<\mathrm{Dec}<13^{\circ}$ sky in the frequency range of
50-350 MHz. The core array with a maximum baseline of 50 km will have
a resolution of $11''$, and its collecting area of 800,000 m$^2$ will
reach a flux density limit of $2\mu \mathrm{Jy}\;
\mathrm{hr}^{-1/2}$. In 10-min integrations, the 600 m core array can
reach 0.05 mJy. It is predicted that SKA1-LOW will find about 7,000
normal pulsars and about 900 millisecond pulsars \citep{kbk+14},
mostly in the closest half of the Galactic halo. SKA1-MID will operate
in 5 different frequency bands; of relevance to this study are the
three bands that overlap with those of SKA1-SUR: band-1: 350--1,050
MHz (MID) or 350--900 MHz (SUR); band-2: 950--1,760 MHz (MID) or
650--1,670 MHz (SUR); and band-3: 1,650--3,050 MHz (MID) or
1,500--4,000 MHz (SUR). Band-1 will be used to survey the halo pulsars
at high Galactic latitudes, band-2 will be used at mid latitudes and
to reach pulsars in the furthest reaches of the Galactic disk, and
band-3 will be used at very low latitudes (e.g.\,$|b|<10^{\circ}$).
According to simulations, SKA1-MID will find about 9,000 normal pulsars
and about 1400 millisecond pulsars (see Fig.\ 3 of \citealp{kbk+14})
distributed in about $2/3$ of the Galactic disk and reaching 5 kpc
farther than the Galactic center. 
With approximately 50\% of the sensitivity of SKA1, the early phase
SKA1 will discover only 1/4 or 1/3 of the above predicted numbers of
pulsars (assuming that similar surveys are done).
SKA2-LOW is expected to discover 11,000 pulsars, including 1500
millisecond pulsars, and SKA2-MID will find 24,000 -- 30,000 pulsars,
of which 2,400 -- 3,000 will be millisecond pulsars. The most distant
pulsars will be about 10 kpc beyond the Galactic center in the other
half of the disk.

The large numbers of pulsars discovered by SKA1 and SKA2 can be used to
constrain the structure of the magnetoionic medium with unprecedented
detail.
Using pulsar dispersion measure (DM) and scattering measure (SM)
estimates, combined with parallax distance estimates, we will
construct the three-dimensional electron density distribution of the
Milky Way, especially the poorly-explored region around the Galactic
center and in the halo, as discussed in Section 2.
Furthermore, the small-scale structure of the magnetoionic medium will
be illuminated by high quality scintillation data and the dynamic
spectra of many bright pulsars.
Pulsar Faraday rotation measure (RM) estimates, combined with pulsar
DM estimates, will be used to reveal the detailed three-dimensional
structure of the Galactic magnetic field, as shown in Section 3.
In fact, the high sensitivity of SKA1 and SKA2 should also
enable the discovery of a large number of pulsars in nearby galaxies,
which opens the window for the study of the intergalactic medium
(IGM), as discussed in Section 4.  The conclusions of this chapter are
presented in Section 5.

\section{SKA1 for electron density distribution}

A better understanding of the electron-density distribution is crucial
for estimating distances to large numbers of distant pulsars and
deriving the magnetic field from Faraday rotation measurements (see
Sect. 3). The ionized gas in the ISM is not uniformly distributed. In
the Galactic disk, high density clumps are associated with
H\,\textsc{ii} regions and bright stars, such that the distribution of
higher density ionized gas follows the spiral arms. There are also
many voids that are associated with bubbles or superbubbles. The
ionized gas can be directly imaged by H$\alpha$ surveys \citep{fin03}, 
but its distribution throughout the Galactic disk is difficult
to ascertain owing to our location near the edge of the Galactic
disk. Measurements of ionized gas in the Galactic halo and
in the area around the Galactic center are limited.

\subsection{Pulsar distances and the electron density distribution}

The dispersion measure (DM) of a pulsar is the integral of the free
electron density between the pulsar and us, $ {\mathrm
  DM}=\int_\mathrm{us}^\mathrm{PSR} n_e\, d l$.
The DMs of a large number of widely distributed pulsars can be used to
construct a model of the electron density distribution, as long as the
distances to these pulsars can be independently measured.  However, pulsar
distances are difficult to measure. Available methods include 1)
measuring annual parallax using Very Long Baseline Interferometry
(VLBI) astrometry, pulsar timing, or even by direct multi-epoch high
resolution optical images; 2) establishing an association between a
pulsar and a supernova remnant (SNR) that has an estimated distance;
3) observing HI absorption and deriving the kinematic distance with
the Galactic rotation curve. Only with an accurate distance can a
pulsar dispersion measure be converted to an average electron density
along the line of sight, and a large number of such sight-lines (of
different length and in different directions) are necessary to
construct a suitably detailed model of the ionized ISM of the Milky
Way.

Currently, the most widely used model for the electron density
distribution in the Milky Way is the NE2001 model (\citealp{cl02}; 
see the yellow background in Fig. \ref{rm_psr_egr}), which has
received over 800 citations and is an essential reference model
for Galactic studies. At the time it was constructed, distance
estimates were available for $\sim$100 pulsars: 19 via parallax (13
via VLBI, 5 via timing and 1 via optical imaging), 74 via HI
absorption, 8 via SNR associations, 16 via globular cluster
associations, and 8 pulsars in the Magellanic clouds.

In order to improve our understanding of the ionized ISM, it is
crucial to accurately estimate the distance to as many pulsars as
possible. Since NE2001 was finalized, a wealth of new information on
pulsar distances has been obtained, with the number of measured
parallaxes\footnote{see e.g.\,
  http://www.astro.cornell.edu/research/parallax/} increasing
five-fold. In particular, the number of precise measurements made with
Very Long Baseline Interferometry has exploded, predominantly due to
careful work with the Very Long Baseline Array \citep{bbgt02,cbv+09,dbc+11}. 
Using pulsars at known
distances, which come from either parallax measurements or by
association with Galactic globular clusters or the Magellanic Clouds,
\citet{gmc+08} found that the ionized thick
disk in the Milky Way could have a scale height of
$1.8^{+0.1}_{-0.2}$~kpc, which is almost a factor of two larger than
the scale height used in NE2001 and its predecessor, the TC93 model by
\citet{tc93}. By also including other model
components from TC93 and NE2001 when fitting for the scale height of
the thick disk, \citet{sch12} subsequently showed
that the difference in scale height is probably not so large. These
results demonstrate that measuring geometric parallaxes for a large
sample of pulsars has the potential to considerably change our
understanding of the structure of the ionized Milky Way. A revised
"NE2014" model incorporating the latest distance and scattering
constraints would already form a considerable advance in our knowledge
of the ionized ISM.

However, despite the recent advances in VLBI precision (where parallax
errors of 0.02 mas and better are now attainable; see e.g.\,\citealp{dbl+13}), 
most pulsars with a geometric distance
measurement are still in the local Galactic neighbourhood; the
existing VLBI and timing parallaxes span 6.4 to 0.2 mas, and the
median of 64 distances measured by parallax is just $\sim$0.9 kpc. To
make a truly Galactic-scale model, many new constraints at larger
distances are required.

Fortunately, in addition to discovering many new pulsars, the SKA will
also excel in the measurement of their distances.  Precision timing of
millisecond pulsars with SKA1 will yield a significant number of
timing parallax distances widely distributed across the Galaxy \citep{kbk+14}. 
The exact number will depend on the observing time
allocated to pulsar timing programs, but observations for pulsar
timing arrays (\citealp{j+14} in this volume) will ensure a
moderately sized sample.

SKA1-VLBI observations (in which SKA1-mid and SKA1-survey are used as
sensitive phased-array elements in a VLBI array; \citealp{pgr+14})
will provide even greater parallax precision, capable of measuring
distances across the Galaxy.  This technique will also be more widely
applicable because all radio pulsars can be targeted, not only
millisecond pulsars.  Observing time will likely limit the number of
sources for which VLBI parallaxes can be obtained with SKA1, but it
will remain possible to obtain a sample of sources that cover a
representative region of the Galaxy.  Both timing and VLBI parallaxes
will be possible with early-phase SKA1 at 50\% sensitivity, although
in each case they will be unable to access the faintest members of the
population.

With SKA2, the situation improves further.  The increase in collecting
area will make even higher precision timing parallaxes possible for
fainter sources; for imaging observations, the addition of several
thousand km baselines to the SKA will make VLBI-style observations
possible with SKA2 alone.  \citet{stw+11} estimated that of order
10,000 pulsar distances could be measured by SKA2, with the majority
obtained via imaging observations.

\subsection{Wide-band observations of pulsar scattering}

The emission region of pulsars is very small (less than a few hundred
kilometers), such that pulsars are effectively point radio sources (with
angular sizes of nanoarcseconds). When pulsed radio emission from a
point source passes through the inhomogeneous interstellar plasma, two
manifestations of multi-path propagation effects can be
observed: pulse broadening in the time domain and a scattering disk in
the image domain. With the SKA, the inhomogeneous interstellar plasma will
produce observable effects on lines of sight to thousands of
pulsars. These effects will be most significant in the SKA1-LOW band,
and combination with the wide frequency range in the two lower bands
of SKA1-MID and SKA1-SUR will enable the unprecedented use of
interstellar scattering to study the properties of the intervening
medium.

\begin{figure}
\centering
\includegraphics[width=85mm]{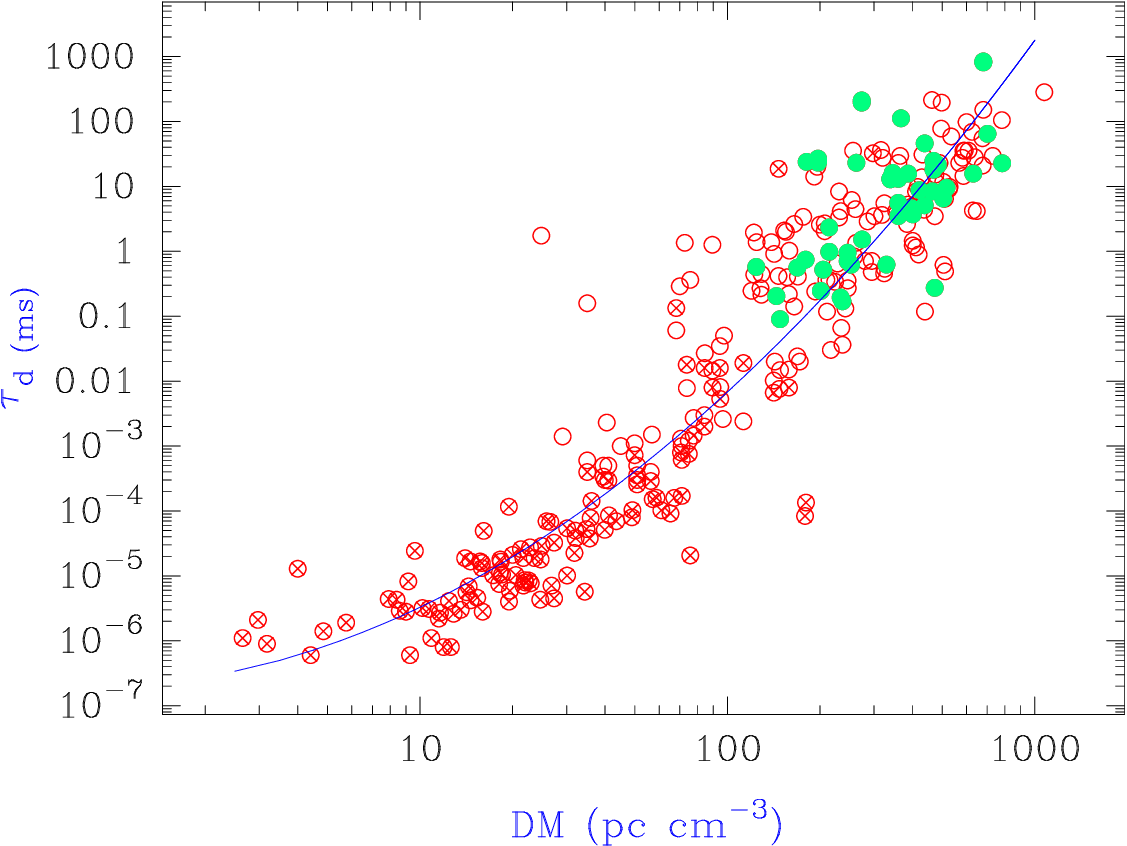} 
\caption{The time scales of pulse broadening are related to pulsar DM,
  in addition to the frequency dependence $\nu^{-\alpha}$ (see \citealp{brc+04}
  for details: the filled dots denote measurements from
  multifrequency profiles presented in that paper, crossed circles
  denote measurements based on pulsar decorrelation bandwidths, and
  open circles denote previous measurements in the literature.  The
  line is the best fit of equation 2.1 to the data).}
\label{tauDM}
\end{figure}

To date, pulse broadening has been measured at several frequencies for
only about 150 pulsars \citep[e.g.][]{lkm+01,lmg+04,brc+04,ldkk13}. 
For nearby pulsars (within a few
kpc) the pulse broadening time scale varies with radio frequency as
$\tau \sim \nu^{-\alpha}$, where $\alpha\sim4.4$, which is consistent
with a Kolmogorov spectrum of electron density fluctuations. For
distant pulsars, the $\alpha$ values deviate from a single thin screen
model. Because ISM irregularities are associated with spiral arms,
the most distant pulsars with sight-lines passing through several spiral arms
should exhibit very different scattering behaviour that is not as simple
as the Kolmogorov spectrum \citep{ldkk13}. The time scale
of pulse broadening is related to the pulsar DM by
\begin{equation}
\label{eqn:t_scat}
\log \tau\simeq a + b (\log \mathrm{DM}) + 
c (\log \mathrm{DM})^2 -\alpha \log \nu
\end{equation}
(see Fig.2 of \citealp{brc+04}) with $a=-6.46$, $b=0.154$ and
$c=1.07$. The estimated value of $\alpha=3.86\pm0.16$ is
significantly lower than the value of 4.4 that is appropriate for a
Kolmogorov medium. However, the dispersion of $\tau$ estimates spans
one or two orders of magnitude in Fig.~\ref{tauDM}. Some pulsars with
high DMs do not show strong scattering as expected, such as PSR
B2002+31 \citep{l71} and the Galactic center pulsar PSR J1745$-$2900
\citep{sle+14}. This probably indicates a predominantly
uniform medium along these lines of sight. It is possible that the
interstellar plasmas in the Galactic halo and near the Galactic center
have fewer irregularities and are more uniformly distributed than in the
disk. 

The scattering disks of only a few pulsars have been observed to date
\citep[e.g.][]{laz04}. The angular diameter of the disk depends on radio
frequency as $\theta\sim\nu^{-\beta}$, where $\beta=-2.2$ for an
electron density distribution described by a Kolmogorov spectrum.  
Note that angular broadening measurements often indicate
$\beta\sim2.0$ (e.g.\,at the Galactic Centre) and decorrelation
bandwidths often scale as $\nu^{-4}$, both of which indicate that the
diffractive scale probed by pulsar observations is within the inner
scale of the turbulent cascade in the ISM.
SKA-VLBI \citep{pgr+14} will provide the sensitivity, wide
frequency coverage, high resolution, and high dynamic range required
to overcome the technical challenges associated with observations of
angular broadening.

Scattering in the ISM operates as an interferometer whose maximum
baseline is of order the size of the scattering disk, up to tens of AU
\citep[e.g.][]{pl14,pmdb14}. Pulsar longitude-resolved
SKA-VLBI observations at low frequencies can constrain source
structure on nanoarcsecond scales, which will enable us to image
pulsar magnetospheres on scales down to $\sim$10~km; for more details,
see \citet{kja+14} in this volume.

\subsection{Wide-band observations of pulsar scintillation}

Owing to the inhomogeneity of the ISM and the high relative velocities
of neutron stars, pulsars exhibit flux density modulations as a
function of radio frequency and time, known as scintillation.  To
characterize the diffractive scintillation, dynamic scintillation
spectra are analyzed using a two-dimensional autocorrelation function
to obtain the temporal and spectral scales of decorrelation, which are
related to the diameter of scattering disk.
Small-scale structures in the ISM cause rapid intensity variations
(diffractive scintillation) and large-scale density structures cause
slow refractive scintillation.

Very high sensitivity is required to observe the dynamic spectrum of a
pulsar, which must be detectable in a narrow band in a short
integration time.  Accordingly, dynamic spectra have been observed for
only the brightest pulsars \citep[e.g.][]{grl94}.  Owing
primarily to the high velocities of pulsars, the differences between
dynamic spectra observed at different epochs probe the irregularities
of the ISM on scales ranging of $\sim10^8$ to $10^{13}$~m.

Dynamic spectra with high signal-to-noise ratio ($S/N$)
can be further analyzed via the two-dimensional Fourier transformation
to yield the power spectrum of the dynamic spectrum, known as the
secondary spectrum.
Secondary spectra typically exhibit power concentrated in parabolic
arcs, and each point on this arc arises from interference between
points in the scattered image of the source. The parabolic shape
arises from differential Doppler shift and differential delay between
points in the scattered image, which vary linearly and quadratically,
respectively, with scattering angle \citep[e.g.][]{crsc06}.
Features in the secondary spectra also vary as a function of radio
frequency and observing epoch, see e.g.\,figures~1, 2 and 3 in 
\citet{crsc06} and figure 1 of \citet{sti06}. Multiple parabolic
arcs are likely produced by multiple scattering screens along the line
of sight.

The best observations of secondary spectra have been carried out at Arecibo
\citep[e.g.][]{crsc06}. Using the core array of SKA1-LOW and
SKA1-MID, high $S/N$ dynamic and
secondary scintillation spectra can be obtained for many pulsars in a
wide-range of low frequencies; these can be used to study both
the static and dynamic structure of the ISM.

\subsection{Interstellar weather} 

The large collecting area and large fractional bandwidth at low
frequencies of SKA-LOW combined with bands 1 and 2 of SKA-MID or
SKA-SUR (including SKA1 and the early phase of SKA1) will provide
very accurate DM and RM estimates for thousands of known and newly
discovered pulsars.
Owing to the $\lambda^2$ dependence of dispersive and refractive
effects in the ISM, DM and RM estimates are best made at lower
frequencies using wider bandwidths, as long as the pulse profile is
not adversely affected by scattering.

Monitoring DM variations over time probes the structure of the ionized
gas on length scales of the order of $10^{8}$ -- $10^{13}$~m.  The
time scale for DM variations depends on the velocity of the pulsar and
the characteristic size of clouds or filaments of ionized gas along
the line of sight.   Increased sensitivity
enables the detection of DM variations on short time intervals due to
small-scale clouds.  
By monitoring short-term DM variations along large numbers of pulsar
lines of sight, SKA1 will probe the physical processes acting within
the interstellar plasma that generate or maintain sub-AU density
fluctuations.

Monitoring long-term DM variations reveals large clumps of ionized gas
drifting across the lines of sight.
Such observations have been made at Parkes using timing array pulsars
over several years \citep{yhc+07,kcs+13} and a sample
of 168 young pulsars over 6 years \citep{pkj+13}.  DM and RM
variations of pulsars near the ecliptic plane can also be used to
study the solar wind \citep{ychm12}.
Note, however, that correcting ionospheric Faraday rotation is
essential for such experiments \citep[e.g.][]{ssh+13}.

\begin{figure}
\centering
\includegraphics[angle=270,width=125mm]{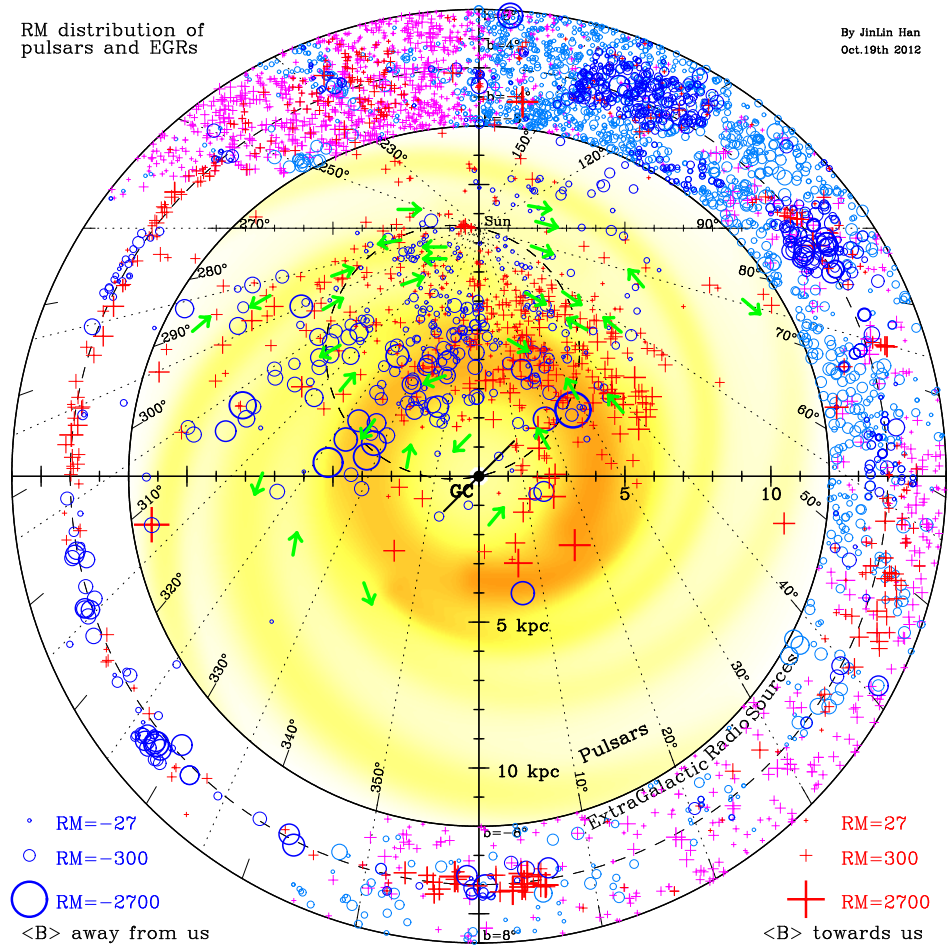}
\caption{The RM distribution of 736 pulsars located within $|b|<8^{\circ}$
  projected onto the Galactic plane. The background shows the
  approximate locations of spiral arms used in the NE2001 electron
  density model \citep{cl02}. RMs of extragalactic radio
  sources \citep{xh14b} located within $|b|<8^{\circ}$ are displayed in the
  outer ring according to their $l$ and $b$, with the same convention
  of RM symbols and limits. The derived large-scale structure of
  magnetic fields in the Galactic disk are indicated by arrows.  See
  \citet{han13} for details.}
\label{rm_psr_egr}
\end{figure}

\section{Interstellar magnetic fields}

Pulsars are ideal probes of interstellar magnetic fields for the
following four reasons: 1) owing to the high degree of linear
polarization of pulsar signals, the RM is easily measured; 2)
intrinsic Faraday rotation in the pulsar magnetosphere is negligible,
such that the observed Faraday rotation is completely due to the
interstellar magnetoionic medium (after correcting for ionospheric
Faraday rotation); 3) the pulsar DM provides the integrated free electron
column density, allowing the line-of-sight magnetic field to be
decoupled (Eq. 3.1); and 4) there are a large number of pulsars
distributed throughout the Galaxy in both the disk and the halo,
facilitating a three-dimensional picture of the Galactic electron
density and magnetic field. For a pulsar at distance $D$ (in pc), the
RM is given by
$ 
\mathrm{RM} = 0.810 \int_{0}^{D} n_e\, {\bf B} \cdot d{\bf l},
$ 
(in rad~m$^{-2}$). With the pulsar dispersion measure,
$ 
\mathrm{DM}=\int_{0}^{D}\,  n_e\,  d l, 
$ 
(in pc~cm$^{-3}$), we obtain a direct estimate of the magnetic field
strength parallel to the line of sight, $\langle B_{||} \rangle$ in
$\mu$G, weighted by the local free electron density:
\begin{equation}\label{eq_B}
\langle B_{||} \rangle  = \frac{\int_{0}^{D} n_e\,  {\bf B} \cdot d{\bf
l} }{\int_{0}^{D}\,  n_e\,  d l } = 1.232 \;  \frac{\mathrm{RM}}{\mathrm{DM}}
.
\label{eq-B}
\end{equation}
When RM and DM data are available for multiple pulsars along similar
lines of sight, e.g.\,one pulsar at $D_0$ and one at $D_1$, the variation
of DM and RM with distance can be used to derive the field direction
and field strength in the region between $D_0$ and $D_1$,
\begin{equation}
\langle B_{||}\rangle_{D_1-D_0} = 1.232 \frac{\Delta\mathrm{RM}}{\Delta\mathrm{DM}},
\label{delta_rm_dm}
\end{equation}
where $\langle B_{||}\rangle_{D_1-D_0}$ is the mean line-of-sight
field component, $\Delta\mathrm{RM} = \mathrm{RM}_{D_1} -
\mathrm{RM}_{D_0}$ and $\Delta\mathrm{DM} =\mathrm{DM}_{D_1} -
\mathrm{DM}_{D_0}$. This derived field strength is not dependent on
the electron density model, though the pulsar distances may have to be
estimated from the electron density model if they have not been
independently measured (e.g.\ by VLBI or pulsar timing).

\begin{figure}
\centering
\includegraphics[angle=-90,width=105mm]{fig4a.ps} \\[3mm]
\includegraphics[width=95mm]{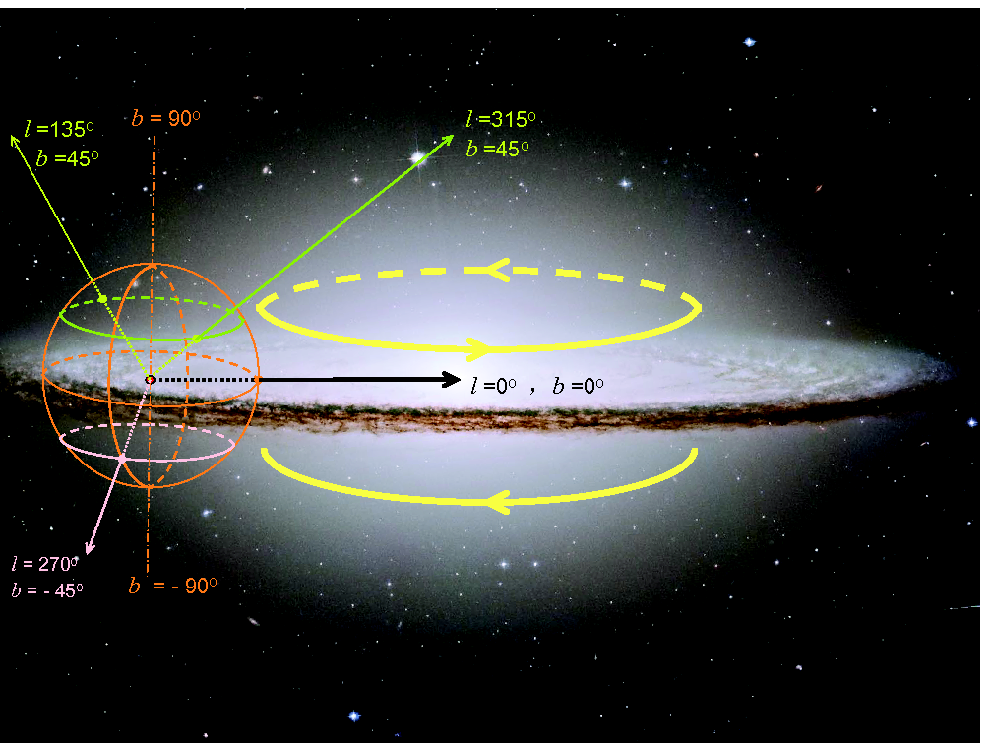} 
\caption{The sky distribution of RMs of extragalactic radio sources
  ({\it top}) and the Galactic coordinates and the azimuthal
  magnetic fields in the Galactic halo ({\it bottom}), see \citet{hmbb97,hmq99}
  and \citet{xh14b}. Currently, pulsar data are
  available only $\sim 3$~kpc from the Sun (see Fig.1).  RMs of
  several thousand distant pulsars to be discovered by the SKA at the high
  latitudes are ultimate data to reveal the large-scale field
  structure in the Galactic halo.}
\label{haloB}
\end{figure}

There are currently over 2,300 pulsars that have been discovered, and
only 30\% of these have published RM values. The RMs in the Galactic
disk have been used to reveal the magnetic field structure in the
closest half of the disk \citep{hml+06}. Over large scales, the
magnetic fields follow the spiral arms; however, many field reversals
are also observed (see Fig.\ref{rm_psr_egr}). It is clear that pulsar
RM data are scarce in many regions of the closest half of the disk;
there are only a few pulsar RM estimates in the most distant regions
of the Galaxy.  The distribution of RMs of extragalactic radio
  sources behind the Galactic disk (see Fig.\ref{rm_psr_egr}) can be
  used to derive the field structures beyond pulsars, and also can 
  constrain the disk-field model \citep[e.g.][]{srwe08}.

The detailed structure of magnetic fields in the Galactic disk will be
revealed by SKA1, which will discover several thousand pulsars in the
closest half of the disk and measure their DMs and RMs, and
SKA2, which will discover at least ten thousand pulsars in the most
distant half of the disk \citep[see][]{kbk+14}. 
Full-polarization observations of the newly discovered pulsars using
SKA1-LOW (50 --350 MHz), and band 1 of SKA1-MID (350 --850 MHz) and
SKA1-SUR (650-- 1,150 MHz) with high spectral resolution (up to
256,000 frequency channels in each band) will determine RMs with
unprecedented precision ($\sim 0.1$ rad~m$^{-2}$) and constrain the
detailed structure of the Galactic magnetoionic medium. Note, however,
that such RM precision is only achievable if the ionospheric
contribution is accurately corrected.

SKA1 and SKA2 will also complement the pulsar RM data by increasing
the density of extragalactic radio sources with RM estimates.
Some constraints on the large-scale structure of the Galactic magnetic
field have been derived from the RMs of extragalactic radio sources,
namely the striking antisymmetry in the inner Galaxy and the
large-scale toroidal magnetic fields in the Galactic halo with
reversed directions above and below the Galactic plane (see
Fig.\ref{haloB}). These structures were originally proposed by 
\citet{hmbb97,hmq99} and later modeled by other authors 
\citep[e.g.][]{ps03,srwe08,ft14}. However, the detailed properties of 
the halo magnetic field
(e.g.\ the variation of field strength with radius and height) are not
yet well constrained. Thousands of distant pulsars in the Galactic
halo will be discovered by the SKA and their RMs will reveal the
large-scale magnetic field structure in the Galactic halo.

In addition to the large-scale structure of the Galactic magnetic
field, a variety of experiments probe small-scale structures in the
interstellar magnetoionic medium.
For example, compared to foreground pulsars along similar lines of
sight, pulsars behind H\,\textsc{ii} regions have significantly
different Faraday rotation measures \citep[e.g.][]{mwkj03}.
Evidence of similar small-scale fields associated with H\,\textsc{ii}
regions, supernova remnants and filaments has also been found in the
RMs of extragalactic radio sources \citep[e.g.][]{hmg11,srw+11,ssf13}. 
Such regions may be explored
in greater detail by exploiting the increased density of extragalactic
radio sources detected by SKA2, SKA1, and the early phase of SKA1.
See \citet{ha+14} for details.

Clearly, interstellar magnetic fields exist over a broad range of
spatial scales, from large Galactic scales to very small dissipative
scales. Determination of the magnetic energy spectrum offers a solid
observational test for dynamo and other theories of Galactic magnetic
field origin.
The magnetic energy spectrum is currently constrained for only a small
range of wavenumbers in relatively few regions.  Magnetic fields on
small spatial scales should follow the the Kolmogorov spectrum. 
\citet{ms96} found that structure functions of rotation measure
and emission measure were consistent with a Kolmogorov spectrum of
three-dimensional turbulence in magnetic fields up to 4 pc, but with
two-dimensional turbulence between 4 pc and 80 pc.
Using pulsar RMs in a large region of the Galactic disk,
\citet{hfm04} obtained a power law distribution of magnetic field
fluctuations described by $E_B(k)= C \ (k / {\rm kpc^{-1}})^{-0.37\pm0.10}$ 
over spatial scales from $1/k=$ 0.5~kpc to 15~kpc.
An apparent turn over in the spectrum between 0.5 kpc and 80 pc
is not yet clear \citep{han09}.
In combination with the RM spectra of intervening polarised sources,
the dense grid of pulsars discovered with the SKA will probe the
energy spectrum at least down to ~100 pc scales.

\section{The extragalactic magnetoionic medium}

In addition to significantly improving our understanding of the
physics of the ISM in the Milky Way, the SKA will reveal the electron
density distribution and magnetic field structure in nearby galaxies
and the IGM. First, it will measure the DMs and RMs of extragalactic
pulsars (especially their giant pulses) and single pulses of
extragalactic origin, such as the Fast Radio Bursts 
\citep[FRBs;][]{lbm+07,tsb+13,sch+14}; see \citet{m+14} 
in this volume for more details. Second, the SKA will
yield high resolution observations of megamaser Zeeman splitting in
nearby galaxies \citep[e.g.][]{rqh08,mh13}; for more details, 
see \citet{rob14} in this volume.
Third, the SKA will observe the polarization of all radio sources in
the sky over a wide band, yielding high precision RMs of a large
number of distant radio sources shining through galaxies, clusters of
galaxies and cosmic web in the nearby universe; see \citet{he+14}, 
\citet{J+14} and \citet{bbc+14} in this volume.

The high sensitivity, wide field of view, and wide bandwidth of SKA1
and SKA2 will yield a large sample of pulsars in nearby galaxies such
as M31 and M33 (see \citealp{kbk+14}, this volume).  At present, DMs
of extragalactic objects have been estimated for only a handful of
pulsars in the Magellanic Clouds \citep{mfl+06} and a few
FRBs \citep{tsb+13}. Scattering has been detected and studied
for only four FRBs \citep[e.g.][]{tsb+13}; the RM of an FRB is
yet to be observed.

\begin{figure}
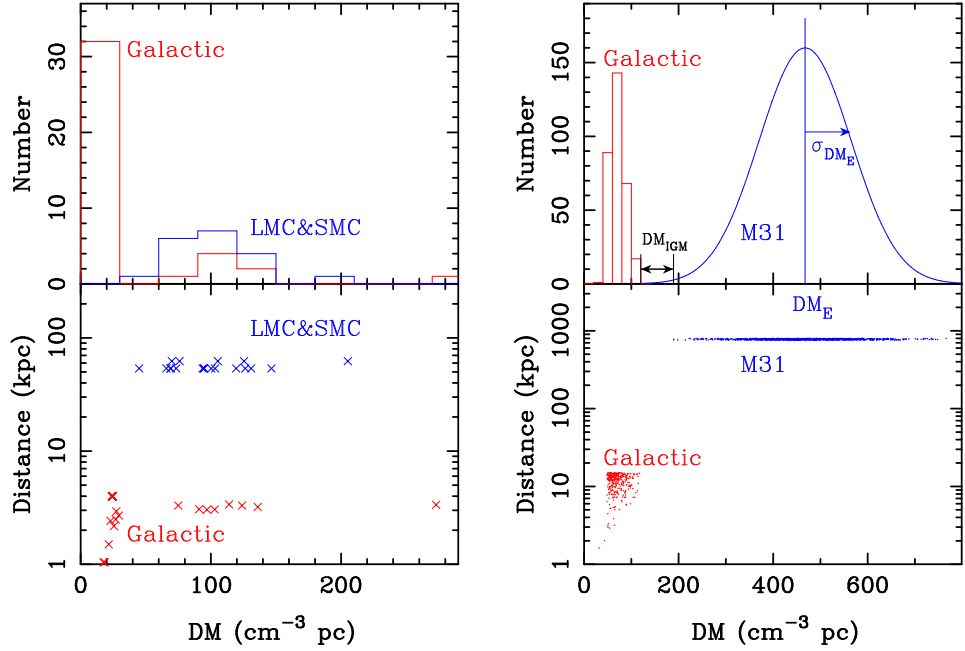

\centering
\includegraphics[angle=270,width=60mm]{fig5a.ps} \hspace{5mm}
\includegraphics[angle=270,width=60mm]{fig5b.ps}
\caption{DM distributions of extragalactic pulsars (blue) and
  Galactic pulsars (red) within a $10^{\circ}$ radius around the host
  galaxy. Current data for pulsars in the large and small Magellanic
  Clouds are compared with Milky Way pulsars {\it on the left} and
  simulated data for pulsars in M31 are compared with simulated Milky
  Way pulsars {\it on the right}. Currently a small sample of pulsars
  in the Magellanic Clouds and a small sample of foreground Galactic
  pulsars around the clouds are not enough to deduce the
  intergalactic DM (or RM). However, the SKA will discover a much larger
  sample of Galactic and extragalactic pulsars (e.g.\ in M31) for this
  purpose. The DM range of extragalactic pulsars depends on the
  inclination angle of the host galaxy disk.
\label{fig:IGM}}
\end{figure}

Two key steps are needed to derive intergalactic DM and RM estimates from
observations of extragalactic pulsars or pulses: 1) subtraction of the
foreground DM or RM contribution from our Milky Way; and 2)
constraining the local contribution from the host galaxy. Both steps
require a large sample of pulsars. Estimating the foreground column
density of electrons in the Milky Way requires a large sample of
Galactic pulsars in the region of sky immediately around the direction
of the host galaxy. To estimate the Galactic foreground contribution
to the measured Faraday rotation also requires a large number of
extragalactic radio sources around the direction of the host galaxy.
SKA1 will observe a large number of extragalactic RMs as part of the
cosmic magnetism project \citep[e.g.][]{J+14}; the
average RM of background radio sources, $\langle
\mathrm{RM}_\mathrm{BGS} \rangle$, in the direction of a host galaxy
represents the foreground contribution of the Milky Way to the
estimated RM of an extragalactic pulsar or pulse \citep[e.g.\,see][]{xh14b}.

For an individual extragalactic pulsar or pulse, it is difficult to
estimate the dispersion and Faraday rotation contributed by the host
galaxy. When a large sample of pulsars are discovered in a host
galaxy, the minimum DM of these pulsars,
$\min(\mathrm{DM}_\mathrm{extraPSRs})$, can be used as the upper limit
of DM$_\mathrm{IGM}$ plus the Galactic DM foreground,
$\max(\mathrm{DM}_\mathrm{GalacPSRs})$, as shown in
Fig.~\ref{fig:IGM}. When averaged over a sample of extragalactic
pulsars in the same host galaxy, the mean RM contributed by the host
is likely to approach zero.
Therefore, to probe the electron density and magnetic field of the
IGM, we can estimate $\mathrm{DM}_\mathrm{IGM}=
\min(\mathrm{DM}_\mathrm{extraPSRs})-\max(\mathrm{DM}_\mathrm{GalacPSRs})$
and $\mathrm{RM}_\mathrm{IGM}= \langle \mathrm{RM}_\mathrm{extraPSRs}
\rangle - \langle \mathrm{RM}_\mathrm{BGS} \rangle$.

In addition to observing extragalactic pulsars and pulses, the SKA
will be able to reveal extragalactic magnetism in a nearby galaxy by
observing the RMs of either a large number of radio sources behind the
galaxy or diffuse emission from the galaxy \citep[see][]{he+14}. 
On the other hand, when a large sample of RMs of quasars
or other objects with known redshifts are available
\citep[e.g.][]{J+14}, the residual RMs (i.e.\, the
values after subtracting the foreground RM from the observed RM
values) can be used to explore the magnetoionic medium in the cosmic
web and intervening clouds \citep[see][]{xh14a}. The distribution and
statistics of residual RMs can reveal weak magnetic fields and their
evolution over cosmological distances much greater than most
detectable extragalactic pulsars.

\section{Conclusions}

Pulsars are the best probes of the interstellar magnetoionic medium in
the Milky Way.  By measuring dispersion toward a large number of
pulsars and their distances, we can construct a detailed electron
density distribution model.  Scattering measurements derived from
either the extended tails of pulsar profiles at low frequencies or
from pulsar dynamic spectra will reveal the dynamics of small-scale
structures in the ionized ISM.  Faraday rotation probes the large and
small scale structure of interstellar magnetic fields. Following the
discussion in previous sections, we conclude the following:
\begin{itemize}
\item SKA1: Beginning with the early phase of SKA1, wide-band
  polarimetry with SKA1-LOW and SKA1-MID will provides a much better
  understanding of the structure of the interstellar medium. The DMs
  of several thousand newly discovered pulsars spread across the Milky
  Way will dramatically improve the electron density model, especially
  once independent and precise distances can be estimated for a
  substantial fraction of pulsars using both SKA1 pulsar timing and
  SKA1-VLBI. The distribution of pulsar RMs can be used to map out
  magnetic field structures in high detail.
\item SKA2: By performing a complete census of those pulsars in our
  Galaxy that are visible in the southern hemisphere, and in concert
  with large telescopes in the northern hemisphere, SKA2 will complete
  the model of electron density distribution and magnetic field
  structure in {\it the entire Milky Way},
\item Intergalactic medium: SKA1 and SKA2 can be used to discover a
  large number of pulsars in nearby galaxies.  By observing the DMs
  and RMs of a large sample of extragalactic pulsars and subtracting
  the Galactic foreground and host galaxy contributions, we will have
  the unique chance to detect the baryonic content and magnetic field
  of the intergalactic medium. The high sensitivity of the SKA is
  essential to this purpose. In addition, the statistics of RMs of a
  large number of extragalactic radio sources can reveal the weak
  magnetic fields in the cosmic web and other intervening galaxies.
\end{itemize}

\begin{acknowledgments}
JinLin Han is supported by the National Natural Science Foundation
(No.11473034) and by the Strategic Priority Research Program ``The
Emergence of Cosmological Structures'' of the Chinese Academy of
Sciences, Grant No. XDB09010200. The research of Joseph Lazio was
performed at the Jet Propulsion Laboratory, California Institute of
Technology, under a contract with the National Aeronautics and Space
Administration. \\
\end{acknowledgments}


\end{document}